\documentclass{vgtc}                          % final (conference style)
\ifpdf%                                % if we use pdflatex
  \pdfoutput=1\relax                   % create PDFs from pdfLaTeX
  \usepackage{graphicx}                % allow us to embed graphics files
  \DeclareGraphicsExtensions{.pdf,.png,.jpg,.jpeg} % for pdflatex we expect .pdf, .png, or .jpg files
\else%                                 % else we use pure latex
  \usepackage{graphicx}                % allow us to embed graphics files
  \DeclareGraphicsExtensions{.eps}     % for pure latex we expect eps files
\fi%

\graphicspath{{figures/}{pictures/}{images/}{./}} % where to search for the images

\usepackage{color}
\usepackage[usenames,dvipsnames]{xcolor}

\usepackage{microtype}                 % use micro-typography (slightly more compact, better to read)
\PassOptionsToPackage{warn}{textcomp}  % to address font issues with \textrightarrow
\usepackage{textcomp}                  % use better special symbols
\usepackage{mathptmx}                  % use matching math font
\usepackage{times}                     % we use Times as the main font
\usepackage{cite}                      % needed to automatically sort the references
\usepackage{tabu}                      % only used for the table example
\usepackage{booktabs}                  % only used for the table example
\onlineid{0}

\vgtccategory{Research}

\vgtcinsertpkg

\title{ProvThreads: Analytic Provenance Visualization and Segmentation}

  \author{Sina Mohseni, Alyssa Pena, Eric D. Ragan\thanks{e-mail: sina.mohseni, mupena17, eragan@tamu.edu}\\ %
        \scriptsize Texas A\&M University %
 }

\abstract{
Our work aims to generate visualizations to enable meta-analysis of analytic provenance and aid better understanding of analysts' strategies during exploratory text analysis.
We introduce \textit{ProvThreads}, a visual analytics approach that incorporates interactive topic modeling outcomes to illustrate relationships between user interactions and the data topics under investigation.
\textit{ProvThreads} uses a series of continuous analysis paths called \textit{topic threads} to demonstrate both topic coverage and the progression of an investigation over time.
As an analyst interacts with different pieces of data during the analysis, interactions are logged and used to track user interests in topics over time.
A line chart shows different amounts of interest in multiple topics over the duration of the analysis.
We discuss how different configurations of \textit{ProvThreads} can be used to reveal changes in focus throughout an analysis.
}

\CCScatlist{ 
  \category{H.5.m}{Information interfaces and presentation}
}

\begin{document}

\firstsection{Introduction}

\maketitle
Visual analytics tools assist analysts with exploration of large amounts of data to identify, understand, and connect pieces of information.
\textit{Provenance} for data analysis tracks the history of the analysis, including the progression of findings, interactions, data inspection, and visual state~\cite{ragan2016characterizing}.
Our research is motivated by the need to support meta-analysis of analytic provenance by researchers and designers to better understand analysts' strategies, to improve analysis tools, and to design effective training programs for data analysts.
Analyzing user interactions and data provenance can reveal information about the analysis process, help in understanding how the user makes discoveries, and explain different analysis strategies.

In exploratory data analysis, it can be difficult to keep track of the different thoughts and topics considered during analysis, and analysts do not want to have to interrupt their thinking and work flow to annotate their thought process.
Prior work has shown that interaction history can be highly effective for understanding analysis behaviors (e.g.,~\cite{dou2009recovering,blascheck2016va,gotz2009characterizing}).
However, full interaction logs are often long and verbose.
Methods for summarizing and visualizing provenance are needed to provide a high-level overview that can be understood more quickly and easily.
We aim to summarize the analysis process automatically using only the system logs from user interactions with data analysis software, thus avoiding the need for supplemental comments from the analysts.
Summarization can be done by dividing the analysis process into smaller meaningful segments where each segment represents a stage of the analysis.
% ------------------------------ Design rational -------------------------------

% Our work focuses on visualizing and segmenting provenance of exploratory data analysis, in which the topics of analysis are generally unknown.
% Such analyses are open ended, and key factors of interest may change throughout the investigation, leading to the creation, divergence, and merging of numerous lines of thought.

\section{Method}
In this work, we present a method to segment and visualize analysis history of an exploratory text analysis.
To do so, we first need a set of analytic provenance data, then a method to segment the provenance data into smaller stages, and finally the generation of a visualization.

\begin{figure}[bt]
 \centering % avoid the use of \begin{center}...\end{center} and use \centering instead (more compact)
 \includegraphics[width=\columnwidth]{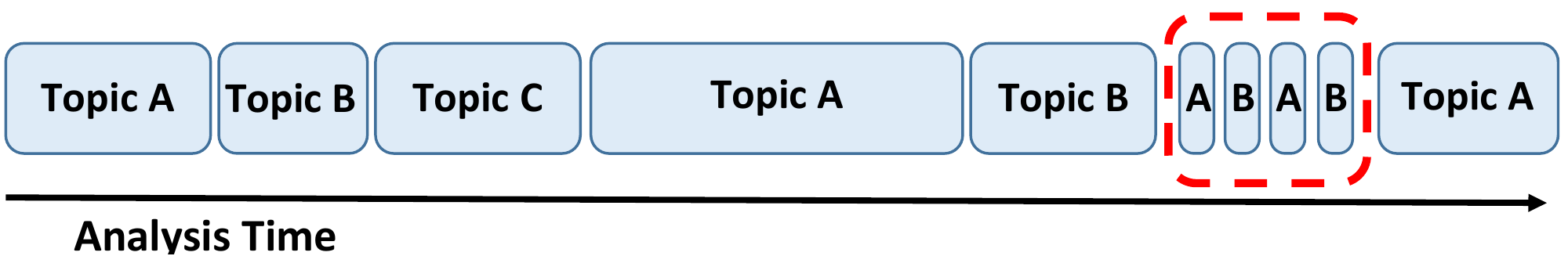}
 \caption{A basic example of how users interact with different topics during an analysis.
 The length of a topic segment indicates the time spent interacting with data related to a particular topic.
 A single long span for one topic corresponds to a focused inspection of one topic, while a burst of multiple short segments (circled in red) might represent a consolidation of topics.
 }
 \label{fig:topicmodels}
\end{figure}

\subsection{Analytic Provenance Data Capture}

To collect provenance test data to design and demonstrate our approach, we conducted a set of user studies using text analysis scenarios from the VAST Challenge datasets (2010 MC \#1, 2011 MC \#3, and 2014 MC \#1).
Three different datasets were used to help assess the robustness and reliability of the design across datasets.
We ran 24 study sessions where participants performed the exploratory analysis. 
To complete the analysis task, participants used a basic visual analysis tool that supports spatial arrangement of articles, the ability to link documents, keyword searching, highlighting, and note-taking.
Anonymized versions of the captured provenance records, user interaction logs, and ProvThreads visualization for all studies are available online  \footnote{https://research.arch.tamu.edu/analytic-provenance/} for research purposes.

\subsection{Analytic Provenance Segmentation}

To segment temporal data, different features and methods can be used to produce meaningful segments, and the selected features largely depends on the nature of the data and the reasons for segmenting.
In our research, we aimed to segment interaction history in a way that corresponds to stages of human analytic thinking.
Through careful review of the captured videos and transcripts of think-aloud comments from the user studies, we studied the times where the participants changed their goals or topics of investigation.
We observed that changes happen when users start looking for new information and connections to support a hypothesis, when they search for new evidence after discovering an insight, or when they continue searching for new clues. 
Topic-change behavior also reveals intuitions about analyst's strategy.
For example, longer periods of time spent on a single topic is indicative of top-down analysis, whereas instances of multiple short, successive topic changes demonstrates bottom-up analysis behavior.
Figure~\ref{fig:topicmodels} shows how topic changes could be used to infer analyst strategy.

Thus, our method uses the interaction history to automatically infer interests and topic changes over the duration of the analysis.
As other researchers have found, identifying user intentions and reasoning form interaction data can be effective~\cite{dou2009recovering,linderresults}, but it is difficult to achieve with concise and accurate representations. 

% \begin{center}
\begin{figure*}[!t]
 \centering % avoid the use of \begin{center}...\end{center} and use \centering instead (more compact)
 \includegraphics[width=5.5in]{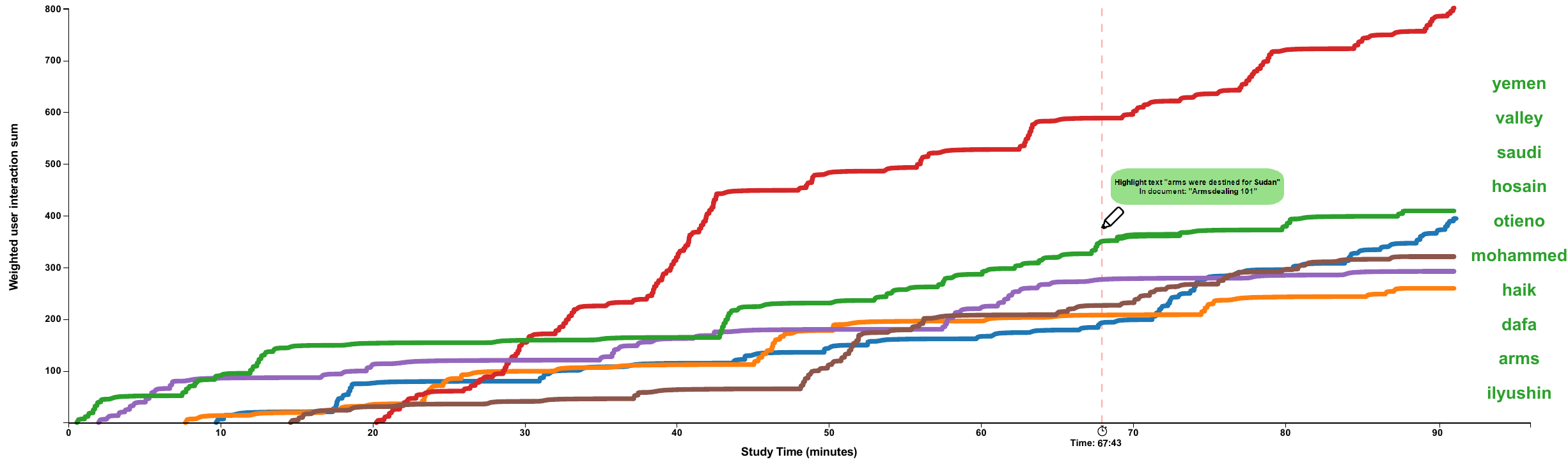} % 0.9\paperwidth
 \label{fig:prov2}
\end{figure*}

\begin{figure*}[!t]
 \centering % avoid the use of \begin{center}...\end{center} and use \centering instead (more compact)
 \includegraphics[width=5.5in]{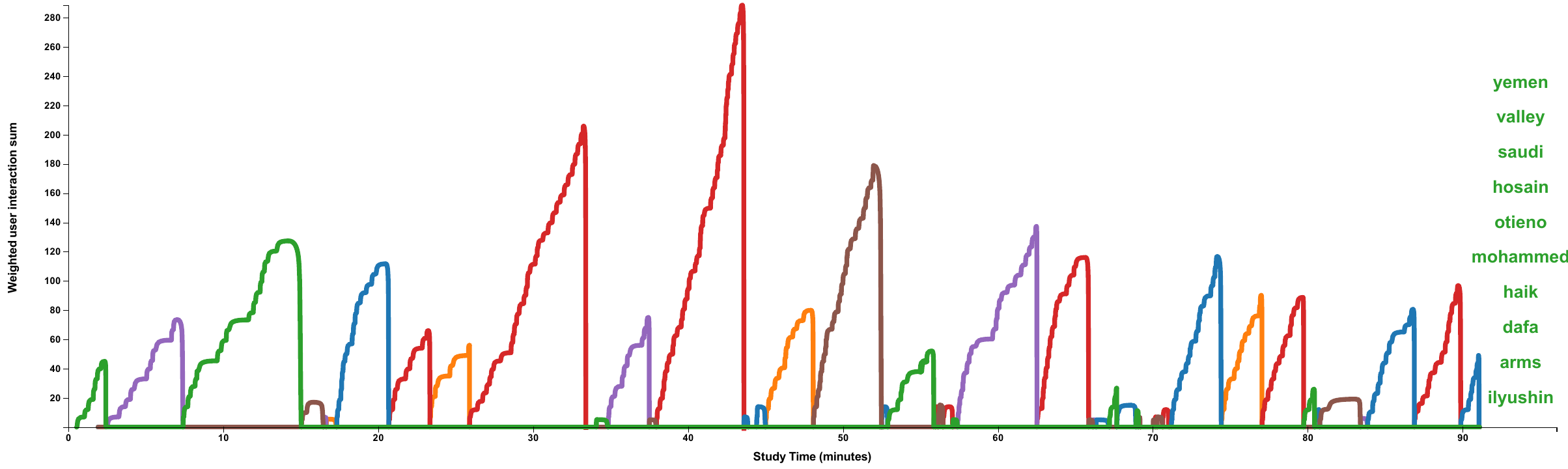} % 0.9\paperwidth
  \caption{
 Two views of the ProvThreads visualization from the same user's interaction logs. Each colored thread corresponds to a topic in the text corpus, and the height increases with additional interactions over time. Brushing over a line shows interaction details in a tooltip, and a list of key terms for the associated topic are shown on the right. Top: The \textit{topic coverage} view shows the accumulation of user focus or inspection in each data topic over the duration of the analysis. 
 Bottom: The \textit{topic segments} view groups topics together based on nearby interactions, and topic height resets when the analyst switches to a new topic group.
 }
 \label{fig:prov3}
\end{figure*}
% \end{center}

\subsection{ProvThreads Method and Design}

ProvThreads is designed to visualize the provenance of topic investigation in a way that connects interaction behaviors with data content.
In the proposed design, analytic provenance segmentation is done by input data classification and assigning user interaction to each data topic.
In our current implementation, we use topic modeling methods to classify text documents using Latent Dirichlet Analysis (LDA) to assign a topic label to each text document.
Although text documents contain a mixture of multiple LDA topics with different
probabilities, the current design labels documents with only the most probable topic  to help simplify the visual summary.
All user interactions from the exploratory text analysis are labeled with the corresponding document's topic.
For instance, if the user is opening a document labeled with topic \textit{A}, writing a note about topic \textit{A}, or searching a keyword from topic \textit{A}, then the associated captured user interaction is labeled with topic \textit{A} as well.

ProvThreads represents the provenance of topic interactions using colored lines that represent different data topics (see Figure~\ref{fig:prov3}).
A user can interactively brush threads to see details (via tool tips) about the interactions or data focus at different times in the analysis.
Additionally, the user can toggle the display of icons the indicate actions for different types of user interactions over time.
The threads design allows for visual identification of patterns of attention to particular topics, and it shows the relationship between interactions and topics.
It shows for how long and how deep the analyst focuses on different pieces of information.
In addition, brushing different topic lines shows the terms associated with each topic (see the word lists on the right side of Figure~\ref{fig:prov3}).
% In addition, the tool supports different filters and parameters to provide a clear view of the topic threads.

ProvThreads supports two visual designs for showing topic interactions: the \textbf{topic coverage} view and the \textbf{topic segments} view.
In the \textbf{topic coverage} view, each topic thread will increase in height when the analyst interacts with the corresponding data topic (see Figure~\ref{fig:prov3}, top).
The main purpose of this design is to track and compare amount of total attention (inferred via interaction) to each topic over time.
% The \textbf{interest threads} view uses different design to show topics of interest at each time of analysis (see Figure~\ref{fig:prov3}, bottom).

The \textbf{topic segments} view follows a different design that emphasizes topic changes (see Figure~\ref{fig:prov3}, bottom).
% In this view, while topic lines still increase along with associated user interactions, each increase causes a reduction in the heights of all other topics threads.
In this view, topic lines still increase along with associated user interactions, but a new topic thread starts when the user switches to a new topic.
% The reduction technique results in attention spikes that highlight changes in which topics the analyst is interacting with.
% This design combines same topic user interactions to create meaningful provenance segments.
To achieve this view, we segment the provenance timeline using a recursive algorithm that combines related topics based on concurrent interactions with data associated with multiple topics.
% based onuser interactions with  and build a higher level of user sub-tasks. % get a more clear view of provenance segments.
This view simplifies and highlights the analyst's topic of interest at each moment and shows clear transition points when the focus changes.

% Interest threads exposes new information regarding analysis flow and analyst's strategies as well as generating analytic provenance segments.
% The interest threads design is configured to use a controllable subtraction ratio that allow a user to visualize analysis focus at different levels of details.
% This way, a user can choose to review the analysis process at a low level (highlighting each individual interaction) or observe higher-level trends in analyst focus. 

\section{Conclusion}

We propose a new method and design to segment and visualize analytic provenance that takes advantage of user interactions with data topics to gives a better picture of an analyst's thought process.
% We present two design configurations that can be used to visualize provenance data and highlight topic changes during analysis.
We prepared a publicly available analytic provenance dataset, and in this ongoing research, we are exploring the effectiveness of a human-in-the-loop approach that supports interactive topic merging to assist segmentation.
% We are also evaluating segmentation accuracy of different methods and assessing the utility of different visual designs for understanding analysis behaviors.

%\bibliographystyle{abbrv}
\bibliographystyle{abbrv-doi}

\bibliography{ref}
\end{document}